\documentclass[prl,reprint,longbibliography] {revtex4-1}

\usepackage[utf8]{inputenc}
\DeclareUnicodeCharacter{03B4}{$\delta$}
\DeclareUnicodeCharacter{2009}{-}

\usepackage{amsmath}

\usepackage{times}
\usepackage[colorlinks=true,linkcolor=blue,urlcolor=blue,citecolor=blue]{hyperref}

\def \dif {{d}}

\date{\today}

\begin{document}
	
\title{Conjectures about the ground-state energy of the Lieb-Liniger model at weak repulsion}
\author{Zoran Ristivojevic}
\affiliation{Laboratoire de Physique Th\'{e}orique, Universit\'{e} de Toulouse, CNRS, UPS, 31062 Toulouse, France}

\begin{abstract}
We develop an alternative description to solve the problem of the ground-state energy of the Lieb-Liniger model that describes one-dimensional bosons with contact repulsion. For this integrable model we express the Lieb integral equation in the representation of Chebyshev polynomials. The latter form is convenient to efficiently obtain very precise numerical results in the singular limit of weak interaction. Such highly precise data enable us to use the integer relation algorithm to discover the analytical form of the coefficients in the expansion of the ground-state energy for small values of the interaction parameter. We obtained the first nine terms of the expansion using quite moderate numerical efforts. The detailed knowledge of behavior of the ground-state energy on the interaction immediately leads to exact perturbative results for the excitation spectrum.
\end{abstract}

\maketitle

Since its introduction in 1963, the Lieb-Liniger model of one-dimensional bosons with contact interaction \cite{lieb_exact_1963,lieb_exact_1963b} continues to fascinate the scientific community. This model is remarkable in many respects and, moreover, describes the physics of realistic systems. It could be nowadays directly realized in experiments with cold gases \cite{kinoshita_observation_2004,paredes_tonksgirardeau_2004,meinert_probing_2015}, enabling us to better understand the correlation effects in one-dimensional many-body systems. The attractive case has deep connections with classical two-dimensional systems, in particular with a surface growth described by the Kardar–Parisi–Zhang equation \cite{kardar_dynamic_1986,kardar_replica_1987}. Rather importantly, the Lieb-Liniger model is integrable and admits an exact solution in terms of the Bethe ansatz \cite{lieb_exact_1963,lieb_exact_1963b}. It thus serves as a benchmark for the effective theories which unavoidably contain various levels of approximations \cite{imambekov_one-dimensional_2012,Giamarchi}. The known results for the above model form a cornerstone for the quantum one-dimensional physics of interacting particles \cite{cazalilla_one_2011,guan_fermi_2013}.

Explicit analytical expressions for various physical quantities of interest in integrable models are often difficult to extract from an exact solution and one is typically restricted to studying  special cases. The relevant information about the system's wave function and the corresponding energy of the Lieb-Liniger model is encoded into the, so called, Lieb integral equation \cite{lieb_exact_1963}. Despite its simple form, the ground-state energy is only known in the limiting cases. Using the systematic procedure of  Ref.~\cite{ristivojevic_excitation_2014}, at strong repulsion one can generate a power series expansion to an arbitrary order in the inverse coupling strength. The other limit of weak coupling, on the other hand, is more difficult to treat since it is singular \cite{lieb_exact_1963}. However, the first three terms were known analytically in this limit for a long time \cite{takahashi_validity_1975,popov_theory_1977,tracy_ground_2016}, until a recent work \cite{prolhac_ground_2017} (see also Ref.~\cite{lang_correlations_2018}) which contains conjectures for the analytical form of the first six terms in the ground-state energy and the general structure of the following ones based on the double extrapolation of the numerical solution of the discrete Bethe ansatz equations.

In this paper we study the Lieb-Liniger model at weak repulsion. We develop an algorithm to efficiently solve the Lieb integral equation in this singular limit to  exceptional precision. The latter feature of numerical results enables us to use the integer relation algorithm \cite{ferguson_analysis_1999,borwein_applications_2000} and identify the analytical expressions for the coefficients in the expansion of the ground-state energy at weak coupling. In the thermodynamic limit, where the particle number $N\to \infty$ and the system size $L\to\infty$ but the density $n=N/L$ is fixed, the ground-state energy can be expressed as $E_0=(\hbar^2 n^2 N/2m)e(\gamma)$, where
\begin{widetext}
\begin{align}
e(\gamma)={}&\gamma-\frac{4}{3 \pi} \gamma ^{3/2} +\frac{\pi ^2-6}{6 \pi ^2} \gamma^2 - \frac{4-3 \zeta (3)}{8 \pi ^3}\gamma^{5/2} - \frac{4-3 \zeta (3)}{24 \pi^4} \gamma^3 - \frac{45 \zeta(5)-60 \zeta (3)+32}{1024 \pi ^5} \gamma^{7/2}\notag\\ & - \frac{3 \left[15 \zeta(5)-4 \zeta (3)-6\zeta (3)^2\right]}{2048 \pi ^6}\gamma^4 -\frac{8505\zeta(7)-2520\zeta(5)+4368\zeta(3)-6048\zeta(3)^2 -1024}{786432 \pi^7}\gamma^{9/2}\notag\\
&-\frac{9[273\zeta(7)-120\zeta(5)+16\zeta(3)-120\zeta(3)\zeta(5)]} {131072 \pi ^8}\gamma^5 +O(\gamma^{11/2}).\label{eq:e}
\end{align}
\end{widetext}
In the above formulas, $m$ is the mass of bosons, while $\gamma=c/n$ is the dimensionless Lieb parameter \cite{lieb_exact_1963}. By $c$ we denote the interaction strength (see the precise definition in the Hamiltonian below). The first line of  expression (\ref{eq:e}) fully agrees with the one of Ref.~\cite{prolhac_ground_2017}. However, our numerical precision can be so high that here we give an analytical form for several more terms. The detailed knowledge of the dependence of the ground-state energy on the density (or, equivalently, on $\gamma$) is valuable information as, e.g., it enables one to find the excitation spectrum of the Lieb-Liniger model \cite{petkovic_spectrum_2018}. In this sense, by finding the ground-state energy one automatically reveals the coefficients in the spectrum of elementary excitations as a function of the momentum \cite{petkovic_spectrum_2018}. The  algorithm developed in this work can be also used to study other problems characterized by similar integral equations. In the following, we derive Eq.~(\ref{eq:e}). 

The Lieb-Liniger model of interacting bosons is defined by the Hamiltonian
\begin{align}\label{Horiginal}
H=-\frac{\hbar^2}{2m}\sum_{i=1}^N \frac{\partial^2}{\partial x_i^2}+\frac{\hbar^2 c}{2m}\sum_{i\neq j} \delta(x_i-x_j).
\end{align}
We consider the repulsive case with positive interaction coupling constant$c>0$, in a system with periodic boundary conditions. The Hamiltonian (\ref{Horiginal}) can be diagonalized by the Bethe ansatz \cite{lieb_exact_1963}. Its ground-state energy can be found from the (rescaled) density of Bethe roots $\rho(x,\lambda)$, which is a continuous function that satisfies the Lieb integral equation
\begin{align}\label{f}
\rho(x,\lambda)-\frac{\lambda}{\pi}\int_{-1}^{1} \dif y \frac{\rho(y,\lambda)}{(x-y)^2+\lambda^2}=\frac{1}{2\pi}.
\end{align}
The dimensionless parameters $\gamma$ and $\lambda$ are connected by the normalization condition 
\begin{align}\label{eq:norm}
\gamma \int_{-1}^{1}\dif x \rho(x,\lambda)=\lambda.
\end{align}
The ground-state function $e(\gamma)$ can then be expressed as
\begin{align}\label{eq:edef}
e(\gamma)=\frac{\gamma^3}{\lambda^3}\int_{-1}^{1} \dif x x^2\rho(x,\lambda),
\end{align}
where in the right hand side of the equation one should express the parameter $\lambda$ in terms of $\gamma$ using their connection via Eq.~(\ref{eq:norm}).

The weakly-interacting limit of the model occurs at $\gamma\to 0$ which corresponds to $\lambda\to 0$. In this limit the kernel in the integral equation becomes a $\delta$-function, which leads to unbounded $\rho(x,\lambda)$ \cite{lieb_exact_1963}. The case of finite small coupling is thus particularly complicated for the analytic treatment, however, some progress has been made \cite{popov_theory_1977,tracy_ground_2016}, leading to the first three terms in Eq.~(\ref{eq:e}).

A convenient way to solve the integral equation (\ref{f}) is to expand $\rho(x,\lambda)$ into a set of complete functions on $[-1,1]$ that we take to be Chebyshev polynomials of the first kind, $T_j(x)=\cos(j \arccos x)$ \cite{abramowitz}. We thus assume the form
\begin{align}\label{eq:rhoChebyshevexpansion}
\rho(x,\lambda)=\sum_{j=0}^{M} c_j(\lambda) T_{2j}(x),
\end{align}
where we take only even polynomials, since $\rho$ is an even function of $x$. The upper limit $M$ in Eq.~(\ref{eq:rhoChebyshevexpansion}) is infinity but will in practice be a large number, as we discuss below. One can then analytically evaluate the integral in Eq.~(\ref{f}) and transform the integral equation into a set of linear algebraic equations for the coefficients $c_j(\lambda)$. 

Using the recurrent relations \cite{abramowitz} for Chebyshev polynomials, $T_{j+1}(x)=2x T_j(x)-T_{j-1}(x)$ for integer $j\geq 1$, greatly simplifies the evaluation of the integral  in Eq.~(\ref{f}). Introducing
\begin{subequations}\label{eq:FG}
	\begin{gather}
F_j(x,\lambda)=\frac{\lambda}{\pi}\int_{-1}^{1}\dif y \frac{T_j(y)}{(x-y)^2+\lambda^2},\\
G_j(x,\lambda)=\frac{\lambda}{\pi}\int_{-1}^{1}\dif y \frac{2y\,T_j(y)}{(x-y)^2+\lambda^2},
\end{gather}
\end{subequations}
we find the recurrent relations
\begin{subequations}\label{eq:FGrecurence}
\begin{align}
F_j(x,\lambda)={}&G_{j-1}(x,\lambda)-F_{j-2}(x,\lambda),\\
G_j(x,\lambda)={}&-\frac{4\lambda}{\pi} \frac{1-(-1)^j}{j(j-2)} +4x G_{j-1}(x,\lambda)-G_{j-2}(x,\lambda)\notag\\
&-4(x^2+\lambda^2)F_{j-1}(x,\lambda),
\end{align}
\end{subequations}
for $j\geq 2$. At $j=2$, the seemingly divergent term $\frac{1-(-1)^j}{j(j-2)}$ must actually be set to zero. The functions $F_j(x), G_j(x)$ at $j=0,1$ can be found directly from the definition (\ref{eq:FG}), while for $j\geq2$ they can conveniently be calculated from the recursion relations (\ref{eq:FGrecurence}). We then transform the integral equation (\ref{f}) into 
\begin{align}\label{eq:clambda}
\sum_{j=0}^M c_j(\lambda)\left[T_{2j}(x)-F_{2j}(x,\lambda)\right]=\frac{1}{2\pi}.
\end{align}
The condition (\ref{eq:norm}) now leads to the expression of the Lieb parameter
\begin{align}\label{eq:gammafinal}
\gamma=\frac{\lambda}{\sum_{j=0}^M \frac{2c_j(\lambda)}{1-4j^2}},
\end{align}
which one can use to transform Eq.~(\ref{eq:edef}) into
\begin{align}\label{eq:efinal}
e(\gamma)=\frac{\sum_{j=0}^M \frac{2c_j(\lambda)(3-4j^2)}{16j^4-40j^2+9} } {\left[\sum_{j=0}^M \frac{2c_j(\lambda)}{1-4j^2}\right]^3}.
\end{align}
For $M\to\infty$, the previous three equations represent a different form of the original ones. In particular, Eq.~(\ref{eq:clambda}) is an exact representation of the Lieb integral equation (\ref{f}). Similarly, Eqs.~(\ref{eq:gammafinal}) and (\ref{eq:efinal}) are our  representations for the normalization condition (\ref{eq:norm}) and the energy function (\ref{eq:edef}).

For the purpose of a highly precise numerical evaluation, rather than the initial expressions (\ref{f})-(\ref{eq:edef}), our preferred starting point are Eqs.~(\ref{eq:clambda})--(\ref{eq:efinal}). For a fixed large integer $M$ we can solve Eq.~(\ref{eq:clambda}) at $M+1$ points $x$ where the highest Chebyshev polynomial $T_{2M}(x)$ reaches  its extrema. This occurs at $x_k=\cos\left(\pi k/2M\right)$, where $k=0,1,\ldots,M$. In this way one obtains a set of $M+1$ linear equations to find the coefficients $c_j(\lambda)$. The functions $F_{2j}(x_k,\lambda)$ are obtained efficiently from the recurrent relations (\ref{eq:FGrecurence}). The approximate solution of the integral equation is then given by substituting them into Eq.~(\ref{eq:rhoChebyshevexpansion}), while $\gamma$ and $e(\gamma)$ are  obtained from Eqs.~(\ref{eq:gammafinal}) and (\ref{eq:efinal}). 

To give an example of the efficiency of our method, for $\lambda=1/10$ and $M=100$ one obtains $\gamma$ and $e(\gamma)$ with a relative error of the order of $10^{-36}$ in one second time on a single core of the processor. By increasing the value of $M$ one obtains progressively more precise results, as detailed in Table \ref{table}. The latter feature of the Chebyshev representation is very important since one can always slightly increase $M$ to verify the precision of the results obtained at smaller values of $M$.

For the purpose of obtaining the result (\ref{eq:e}) one needs $\gamma$ and $e(\gamma)$ at very high precision corresponding to, e.g., $M=600$. We calculated the dependence $e(\gamma)$ at small $\gamma$ by evaluating the system of equations for $50$ different values of $\lambda$ from the interval $(1/60,1/10]$, which can be achieved in around one hour on a single core of the processor.  Instead of $\gamma$ and $e(\gamma)$ we find it convenient to study the related quantities
\begin{align}\label{eq:alpha}
\alpha=\frac{\sqrt{\gamma}}{2\pi},\quad \epsilon(\alpha)=\frac{e(\gamma)}{\gamma}.
\end{align}
We then fitted the numerical data with the function $\epsilon(\alpha)=\sum_{j=0}^{49} a_j \alpha^j$, obtaining $a_0=1.0000\ldots$ that satisfies $|a_0-1|\sim 10^{-59}$. We therefore identified the exact value $a_0=1$. Subtracting the latter unity from the numerically evaluated $\epsilon(\alpha)$ we then fitted the obtained data with the function $\sum_{j=1}^{49} a_j \alpha^j$ which yields $a_1=-2.6666\ldots$. It satisfies $|a_1+8/3|\sim 10^{-56}$, enabling us to identify the exact value $a_1=-8/3$. We continued such a procedure and found the remaining seven coefficients numerically and then found their presumed analytical form. For the coefficient in front of $\alpha^8$ we obtained the numerical value $a_8=-0.3604\ldots$ that differs from the exact coefficient $a_8=-9\zeta(3)/32+135\zeta(5)[1+\zeta(3)]/64-2457\zeta(7)/512$ in absolute value by $10^{-47}$. Such highly precise fitting coefficients $a_0,\ldots,a_8$ which had at least $46$ correct digits were sufficient to use the integer relation algorithm \cite{ferguson_analysis_1999,borwein_applications_2000} that recognizes the approximate number as a certain rational combination of basis vectors that are $1$, $\zeta$-functions \cite{prolhac_ground_2017,lang_correlations_2018}, their powers, and  combinations of $\zeta$-functions, contrary to the conjecture of Ref.~\cite{prolhac_ground_2017}. We have tested the validity of the presumably exact value for $a_8$ by solving the set of equations for $M=800$ and 60 values of $\lambda$ from the interval $(1/40,1/100]$. After fitting the polynomial of an order 59 we have obtained the numerical value that differs in the absolute value from the analytical form for $a_8$ by an order of $10^{-57}$. We have therefore no serious doubts that all the conjectured coefficients $a_0,\ldots,a_8$ are exact. They lead to
\begin{align}
\epsilon(\alpha)={}&1-\frac{8}{3}\alpha+\left(\frac{2\pi^2}{3}-4\right)\alpha^2 -[4-3\zeta(3)]\alpha^3\notag\\
&-\left[\frac{8}{3}-2\zeta(3)\right]\alpha^4 -\left[1-\frac{15}{8}\zeta(3)+\frac{45}{32}\zeta(5)\right]\alpha^5 \notag\\
&-\left[\frac{45}{32}\zeta(5)-\frac{3}{8}\zeta(3)-\frac{9}{16}\zeta(3)^2\right]\alpha^6 -\biggl[ \frac{2835}{2048}\zeta(7)\notag\\
&+\frac{91}{128}\zeta(3)-\frac{105}{256}\zeta(5)-\frac{63}{64}\zeta(3)^2-\frac{1}{6}\biggr]\alpha^7\notag\\
&-\left\{\frac{2457}{512}\zeta(7)+\frac{9}{32}\zeta(3)-\frac{135}{64}\zeta(5)[1+\zeta(3)]\right\}\alpha^8\notag\\
&+O(\alpha^9).
\end{align}
Using relations (\ref{eq:alpha}) we then obtain our final result (\ref{eq:e}).

\begin{table}\label{table}
	\caption{Illustration of the efficiency of our algorithm for two values $\lambda=1/10$ (corresponding to $\gamma=3.403\times 10^{-2}$), $\lambda=1/100$ ($\gamma=3.906\times 10^{-4}$), and for several values of $M$. We give the relative error in the evaluated value of $\gamma$, which is of the same order as for $e(\gamma)$. We notice that our implementation was not fully optimized and therefore  the execution times could be shorter.\label{table}}
	\begin{ruledtabular}
		\begin{tabular}{c|cr|cr}
			&\multicolumn{2}{c|}{$\lambda=\frac{1}{10}$}&\multicolumn{2}{c}{$\lambda=\frac{1}{100}$}\\
			$M$ &relative error&time in $[\mathrm{s}]$ &relative error&time
in $[\mathrm{s}]$\\ \hline
			$100$ & $10^{-36\phantom{0}}$ & $1$& $10^{-16}$ & $1$\\
			$200$ & $10^{-65\phantom{0}}$ &$3$ & $10^{-26}$ &$3$ \\
			$300$& $10^{-93\phantom{0}}$ & $9$ & $10^{-35}$ & $10$ \\
			$400$ & $10^{-121}$ & $22$& $10^{-45}$ & $22$  \\
			$600$ & $10^{-177}$ & $76$ & $10^{-62}$ & $77$ \\
			$800$ & $10^{-233}$ &$213$& $10^{-80}$ &$209$\\
		\end{tabular}
	\end{ruledtabular}
\end{table}

The obtained result for the ground-state energy (\ref{eq:e}) enables us to calculate many other important quantities for the Lieb-Liniger model. The Luttinger liquid exponent $K$ controls the decay of various correlation functions. It follows from the exact relation \cite{lieb_exact_1963b}
\begin{align}
K=\frac{\pi}{\sqrt{3e(\gamma)-2\gamma \frac{d e(\gamma)}{d\gamma} +\frac{1}{2} \gamma^2 \frac{d^2e(\gamma)}{d\gamma^2}}}.
\end{align}
The sound velocity can then be directly obtained from the Galilean invariance relation $v=\pi\hbar n/m K$ \cite{haldane_effective_1981}. Interestingly, the information about the spectrum of elementary excitations is also contained in Eq.~(\ref{eq:e}) as a consequence of the integrability of the model \cite{petkovic_spectrum_2018}. For example, the effective  mass of elementary excitations $m^*$ \cite{imambekov_one-dimensional_2012} can be obtained from the relation $m/m^*=(1-\gamma\partial_\gamma)K^{-1/2}$ \cite{ristivojevic_excitation_2014}, while the coefficients of the cubic and quartic term in momentum follow from Eqs.~(12) and (13) of Ref.~\cite{petkovic_spectrum_2018}.

The successful use of the integer relation algorithm to recognize a numerical constant requires its high precision, which grows by increasing the number of basis vectors \cite{borwein_applications_2000}. This is a serious limiting factor in practice. In the present problem we were able to avoid the numerical integration, which is always a source of a numerical error, in the integral equation (\ref{f}) by making use of the derived recurrent relations (\ref{eq:FGrecurence}) and thus easily produce very precise numerical data. Moreover, the precision can be further increased, when necessary, by simply increasing the number of Chebyshev polynomials $M$, as illustrated in Table \ref{table}. The obtained perturbative expansion for the ground-state energy (\ref{eq:e}) at high order in the regime $\gamma\to 0^+$, together with an analogous result in the other limit $\gamma\to\infty$ \cite{ristivojevic_excitation_2014} should serve as a starting point to study the non-perturbative structure of the Lieb-Liniger model \cite{bender_anharmonic_1973,aniceto_primer_2019}.

In conclusion, we have developed an alternative way to obtain the exact analytical perturbative results for the ground-state energy of the Lieb-Liniger model in the analytically complicated regime of weak interaction. First, we found a way to efficiently obtain highly precise numerical data. We then used the integer relation algorithm to recognize the analytical form of the ground-state energy, which is then additionally verified by even more precise numerical evaluations. The obtained result for the ground-state energy actually contains the information about the whole excitation spectrum. Our work shows how  knowledge from the young discipline of experimental mathematics could be successfully used to treat the long-standing problem in quantum physics.

\textit{Note added.}--In the final stage of preparation of this work, a related preprint appeared \cite{marino_exact_2019}. It reports the first eight terms of $e(\gamma)$ that are in full agreement with our result (\ref{eq:e}).  


%

\end{document}